\documentclass[final]{nature}

\usepackage{color}
\usepackage{graphicx}

\title{Active Stabilization of Terahertz Semiconductor Dual-Comb Laser Sources Employing a Phase Locking Technique}

\author{Yiran Zhao$^{1,2,6}$, Ziping Li$^{1,6}$, Kang Zhou$^{1,2,6}$, Xiaoyu Liao$^{1,2}$, Wen Guan$^{1,3}$, Wenjian Wan$^1$, Sijia Yang$^{1,2}$, J. C. Cao$^{1,2}$, Dong Xu$^{1,5}$, Stefano Barbieri$^{4}$, and Hua Li$^{1,2\star}$}

\begin{document}

\linespread{1.5}\selectfont

\maketitle

\begin{affiliations}
 \item Key Laboratory of Terahertz Solid State Technology, Shanghai Institute of Microsystem and Information Technology, Chinese Academy of Sciences, 865 Changning road, Shanghai 200050, China.
 \item Center of Materials Science and Optoelectronics Engineering, University of Chinese Academy of Sciences, Beijing 100049, China.
 \item School of Information Science and Technology, ShanghaiTech University, 393 Middle Huaxia Road, Shanghai 201210, China. 
 \item Institute of Electronics, Microelectronics and Nanotechnology, University Lille, ISEN, CNRS, UMR 8520, 59652 Villeneuve d'Ascq, France. 
 \item SIMIC Advanced Micro Semiconductors Co., Ltd., 888 2$^{\rm{nd}}$ West Huanhu Street, Pudong, Shanghai 201306, China.
  \item These authors contributed equally: Yiran Zhao, Ziping Li, Kang Zhou. \\
  $^{\star}$Corresponding author. E-mail: hua.li@mail.sim.ac.cn.
 
\end{affiliations}

\begin{abstract}

Dual-comb sources with equally spaced and low phase noise frequency lines are of great importance for high resolution spectroscopy and metrology. In the terahertz frequency range, electrically pumped semiconductor quantum cascade lasers (QCLs), demonstrating high output power, good beam quality, and wide frequency coverage, are suitable candidates for frequency comb and dual-comb operation. Although free running terahertz QCLs can be operated as frequency combs and dual-comb sources, the phase noise originated from the carrier frequency and repetition rate instabilities are relatively high, which hinders the high precision applications. For a single laser frequency comb, the repetition rate can be locked using a microwave injection locking and the carrier frequency can be locked to a highly stable source. However, for the locking of two laser combs, four frequencies (two repetition rates and two carrier offset frequencies) should be simultaneously locked; If one only refers to the dual-comb signal, two relative frequencies, i.e., the offset frequency and repetition frequency of one laser against those of the other laser, should be locked. Although the locking techniques that have been successfully used for a single laser comb can be, in principle, applied to a dual-comb laser source, the complete locking considerably complicates the implementation of such a system. Here, we propose a method to stabilize a terahertz QCL dual-comb source by phase locking one of the dual-comb lines to a radio frequency (RF) synthesizer. This technique forces one of the lasers to follow the “tone” of the other one (keeping the sum of the carrier offset frequency difference and repetition frequency difference between the two laser combs as a constant) by exploiting a laser self-detection that avoids the use of an external detector. Although only one dual-comb line is locked (we don’t lock the two repetition rates and/or the carrier offset frequencies), we show that the phase noise of other dual-comb lines close to the phase locked line is significantly reduced. Finally, through the demonstration of this locking technique, we demonstrate that the dual-comb can generate periodic pulses over a 2 $\mu$s time scale, showing that the terahertz QCL comb without a control of the repetition rate can produce pulsed-type waveforms. The demonstrated approach provides a convenient method to actively stabilize terahertz dual-comb laser sources using a traditional phase locking technique, which can be further utilized for fast gas sensing and spectroscopy in the terahertz range.

\end{abstract}

\linespread{1.2}\selectfont

\section{Introduction}

A frequency comb \cite{HolzwarthPRL,UdemNature} consists of a series of coherent and equally spaced lines covering a broad spectrum, which has revolutionized time and frequency metrology by exploiting its highly stable frequency lines and ultrafast optical pulses. Two frequencies can be used to fully describe a frequency comb, i.e., the carrier offset frequency and the repetition rate. Once these two parameters are known, each line of a frequency comb can be accurately defined. Frequency combs can be straightforwardly used for high-precision spectroscopy by deploying a multiheterodyne dual-comb technique \cite{KeilmannOL2004}, where the beating between two frequency combs with slightly different repetition rates is down-converted in the radio frequency (RF) range using a fast detector. In the last two decades, dual-comb techniques in the near infrared and visible wavelength range have experienced significant developments targeting various applications in spectroscopy, distance measurement, imaging, communications, and so on \cite{VillaresNC,CoddingtonOptica,YanLight2017,Kippenberg2018,PicqueNP2019}. 

Due to the importance of frequency combs and dual-comb sources for fundamental research and a wide range of applications, a significant research effort has been devoted to extending the range of operation of frequency combs to other wavelengths. In this respect, the terahertz wave (0.1-10 THz) is attracting a growing interest due to potential applications in communications, security, health science, environmental monitoring, etc \cite{ZhangNM,TonouchiNP}. For a long time, frequency comb and dual-comb sources directly emitting in the terahertz frequency range have been out of reach. The most popular indirect approach to generate terahertz frequency combs is to down-convert near-infrared frequency combs by nonlinear frequency conversion using photomixing in photoconductive antennas or optical rectification in nonlinear crystals \cite{FinneranPRL,YasuiSR}. However, these down-conversion techniques require an external femtosecond laser pump, resulting in a bulky system with, moreover, a weak optical power.

Semiconductor-based quantum cascade lasers (QCLs) \cite{1stTHzQCL}, showing high output power \cite{HighPower}, wide frequency coverage \cite{Octave,5.2THz}, and high beam quality \cite{WanOE}, are ideal candidates for directly generating frequency combs, as well as for dual-comb operations in the terahertz range. Due to the fact that QCLs are electrically pumped, QCL frequency comb sources offer a small size, weight and consumption power (SWAP) \cite{FortierRev2019}, which allow these comb sources to be used for the realization of compact systems, for spectroscopy, imaging, on-chip communications, etc. The simplest way to obtain comb operation with a free running terahertz QCL is to exploit the four-wave mixing properties of the nonlinear QCL gain medium, together with the group velocity dispersion engineering \cite{BurghoffNP,Octave1,ZhouAPL,PiccardoNature2020}. The coherence of a single QCL comb can be improved by locking its repetition rate using either microwave injection techniques or a saturable absorber \cite{OustinovNC,GellieOE,FaistACS2020,SirtoriLPR2020,LiAS}, as well as by phase locking the carrier offset frequency using a highly stable femtosecond laser \cite{BarbieriNP,CappelliNP,ConsolinoNC2019}. These terahertz QCL combs based on active or passive stabilization techniques, have been shown to produce short optical pluses between 1 and 16 ps \cite{Bachmann,BarbieriNP,5ps,WangLPR,LiAS,WangFH2020}. Concerning dual-comb operation, up to now most terahertz QCL dual-comb sources are operated fully in free running without any control of the repetition rate and/or carrier offset frequency \cite{THzDualComb,YangDC2016,SterczewskiOptica2019,LiACSPhoton,Sterczewski2020}. In principle, the previously mentioned stabilization techniques can be applied to a dual-comb laser source. If one only refers to the locking of the dual-comb signal, two relative frequencies, i.e., the offset frequency and repetition frequency of QCL Comb1 against those of Comb2, should be locked simultaneously, which is heavy to implement.

In this work, we report the frequency stabilization of a terahertz QCL dual-comb source emitting around 4.2 THz, obtained by phase locking one of the dual-comb lines using a phase locked loop (PLL). Despite the fact that only one dual-comb line is locked, we find that the frequency stability of the dual-comb lines that are close to the phase locked line is significantly improved. Finally, we demonstrate experimentally that the phase locked dual-comb signal consists of a periodic pulse train in time domain. Although the dual-comb time trace does not provide the output waveform of one single QCL comb, this result indicates that the waveforms emitted by the two QCLs individually are time periodic, with a pulsed-type behavior. The demonstration of the phase locking of terahertz QCL dual-comb sources is an important step towards the development of high spectral resolution and high speed terahertz spectrometers and imagers.

\section{Experimental setup and laser performances}

\begin{figure*}[!t]
	\centering
	\includegraphics[width=0.9\textwidth]{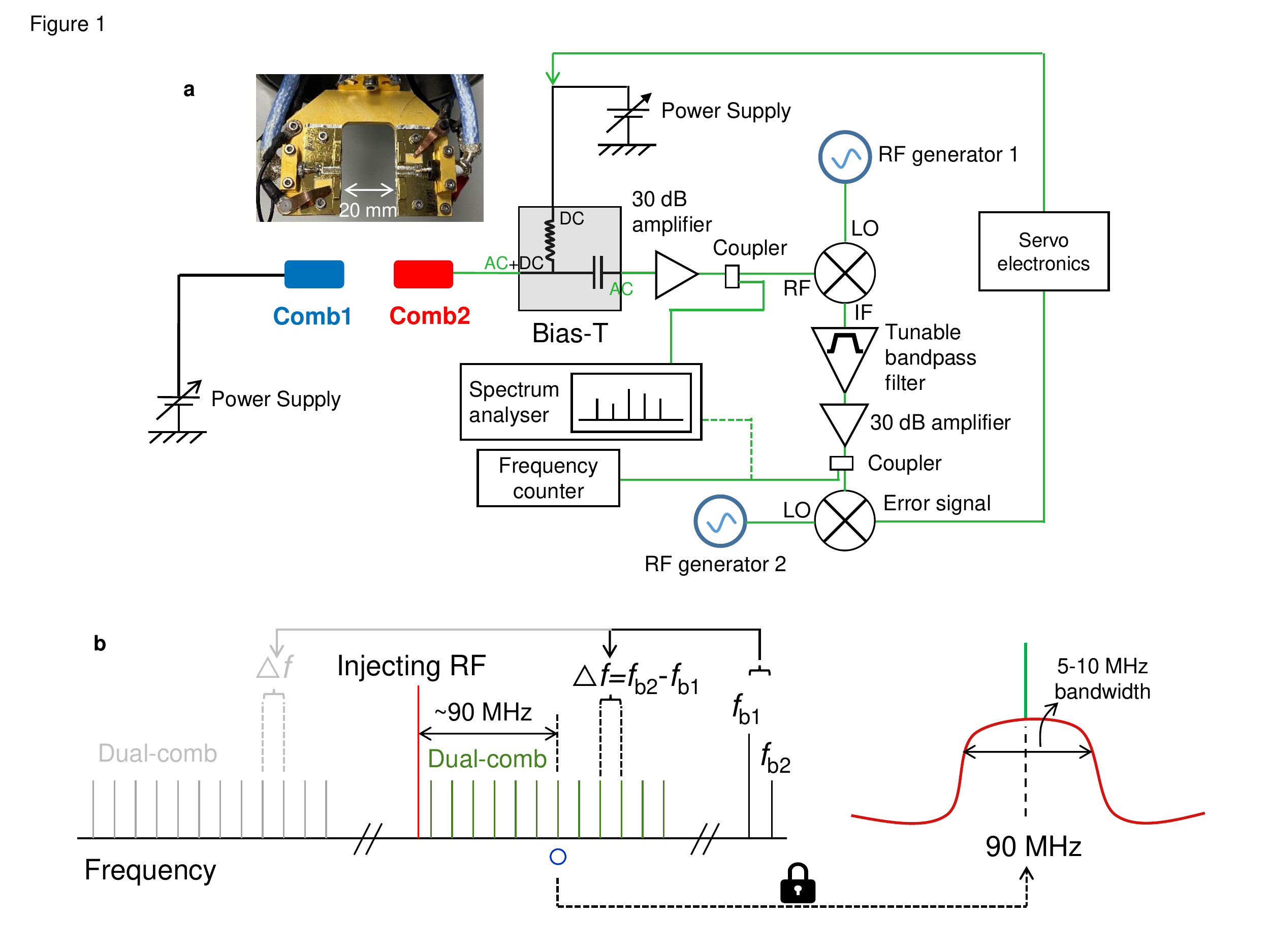}
	\caption{Experimental setup for the phase locking of the terahertz dual-comb laser source. a) Experimental setup. Comb1 and Comb2 are single plasmon terahertz QCLs with a ridge width of 150 $\mu$m and a cavity length of 5.5 mm. RF generator 1 is used to down-convert the multiheterodyne dual-comb signal to 90 MHz. RF generator 2 acts as a local oscillator (LO) to beat with the down-converted dual-comb line for generating the error signal to dynamically control the drive current of Comb2. Green lines represent high speed microwave cables and black lines standard BNC cables. The inset is a photo of the dual-comb lasers mounted on a cold finger: the distance between the two laser front facets is 20 mm. IF: intermediate frequency. b) Frequency synthesis of the dual-comb phase locking experiment. $f$$_{\rm{b1}}$ and $f$$_{\rm{b2}}$ are the fundamental inter-mode beatnote frequencies of Comb1 and Comb2, respectively. $\Delta$$f$ is the difference between $f$$_{\rm{b1}}$ and $f$$_{\rm{b2}}$. The gray and green lines show two groups of dual-comb spectra at the first and second lowest carrier frequencies (see Ref. \cite{LiACSPhoton} for the detailed description of the different dual-comb generation processes). The line spacing of the dual-comb signals is equal to $\Delta$$f$. The injecting RF (red line) is used to down-convert one of the dual-comb lines (marked by the circle) to 90 MHz for the phase locking.}
	\label{setup}
\end{figure*}

{\textbf{Figure \ref{setup}}}a shows the experimental setup of the phase locking of the terahertz dual-comb source. The inter-mode beatnote and multiheterodyne dual-comb signals in the radio frequency (RF) range are measured via Comb2 employing a self-detection scheme (see the Experimental Section for details). In this experiment, only one of the dual-comb lines is finally selected and amplified for the phase locking (see the Experimental Section). The terahertz QCLs used in this work are based on a hybrid active region design (bound-to-continuum transitions for photon emission and fast longitudinal optical phonon scattering for depopulation in the lower laser state) \cite{ScalariHybrid,BTC-RP,WanSR}, which has been experimentally proved to be suitable for frequency comb and dual-comb operation at frequencies around 4.2 THz \cite{ZhouAPL,LiAS}. The nominal cavity length and ridge width for both laser combs are 5.5 mm and 150 $\mu$m, respectively. However, due to the imperfections in the material growth and cleaving processes, the refractive index and the cavity length are not exactly identical for both combs which finally result in a slight difference in the inter-mode beatnote frequencies, i.e., $f$$_{\rm{b1}}$ and $f$$_{\rm{b2}}$. This frequency difference is crucial for the dual-comb generation. As shown in Figure \ref{setup}b, the spacing of the dual-comb spectra is equal to the difference between $f$$_{\rm{b1}}$ and $f$$_{\rm{b2}}$: if the two inter-mode beatnote frequencies are identical, no dual-comb spectrum can be observed. Note that multiple dual-comb spectra at different carrier frequencies in the microwave frequency range can be obtained through different beating processes of the terahertz modes. Not only the beatings of the neighboring terahertz modes of the two QCL combs, but also the beatings between modes that are far separated can generate dual-comb signals with an identical dual-comb line spacing. The detailed description of the different dual-comb generation processes can be found in Ref. \cite{LiACSPhoton}. In Figure \ref{setup}b, we schematically show two dual-comb spectra at the first and second lowest carrier frequencies (gray and green lines).

For the locking of the dual-comb source, we select one dual-comb line, schematically marked by the circle in Figure \ref{setup}b, for the phase locking. First, the entire dual-comb spectrum centered at 4 GHz is down-converted to lower frequencies, around 90 MHz, by beating it with a microwave signal (red line) that is 90 MHz away from the selected dual-comb line. Then, the line at 90 MHz is selected using a bandpass filter with a bandwidth of 5 MHz. Once the phase locked loop is switched on, the line is phase locked by dynamically controlling the drive current of Comb2. It is worth noting that neither the carrier offset frequencies nor the repetition frequencies of Comb1 and Comb2 are locked in this study. The PLL forces Comb2 to follow Comb1 by controlling the dual-comb beating. Although only one dual-comb line is locked using the PLL, we find that the remaining dual-comb lines are significantly more stable compared to the lines without PLL (see {\textbf{Figure \ref{maxhold}}}).

\begin{figure*}[!t]
	\centering
	\includegraphics[width=0.9\textwidth]{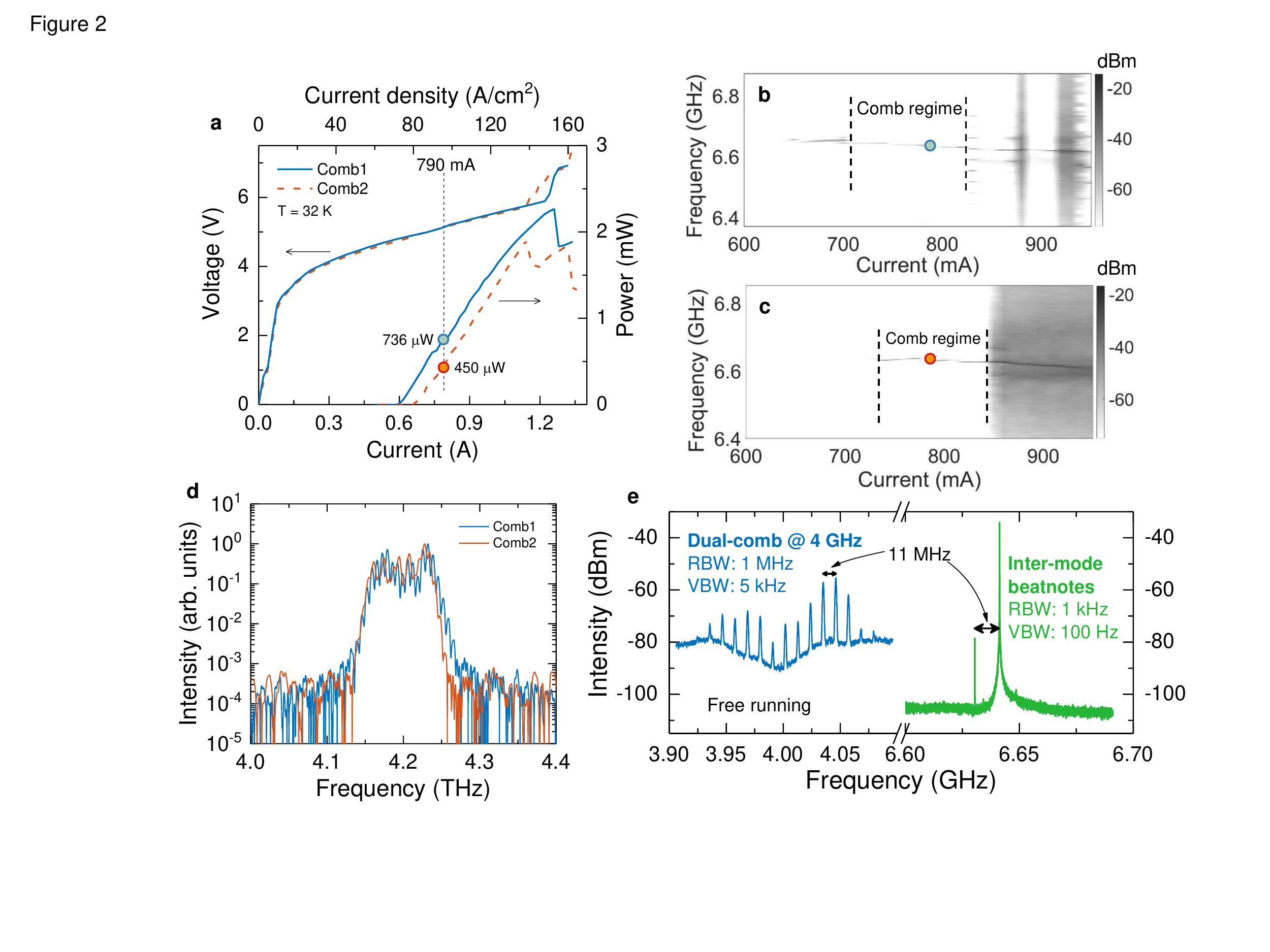}
	\caption{Laser performance and comb characteristics of both lasers with a cavity length of 5.5 mm and a ridge width of 150 $\mu$m. a) Light-current-voltage characteristics of the two QCLs measured in cw mode. The vertical dashed line indicates the drive current of 790 mA used for both lasers for dual-comb operation. b, c) Inter-mode beatnote maps of Comb1 and Comb2, respectively, measured with a RBW of 1 kHz and a video bandwidth (VBW) of 1 kHz. The current step is 10 mA. d) Normalized terahertz emission spectra of Comb1 and Comb2 measured at a drive current of 790 mA. e) Inter-mode beatnote and dual-comb spectra at 4 GHz measured from the laser dual-comb source. For all measurements, the temperature is stabilized at 32 K {\color{red}{with a stability accuracy of 50 mK.}}}
	\label{fr}
\end{figure*}

The measured light-current-voltage curves of Comb1 and Comb2 are shown in {\textbf{Figure \ref{fr}}}a in continuous wave (cw) mode at a stabilized heat sink temperature of 32 K. Overall, the two lasers show uniform electrical and optical characteristics. The measured maximum output powers for the two lasers are slightly different, i.e., 2.2 mW for Comb1 and 1.9 mW for Comb2. This small difference is due to the nonuniform material growth, imperfect fabrication and device mounting processes. Figures \ref{fr}b and \ref{fr}c show the inter-mode beatnote maps for Comb1 and Comb2, respectively. We can see that as the current goes beyond 850 mA, both lasers cannot work as ideal frequency combs since the inter-mode beatnote measured in this current range is either broad or multi-line structured. To obtain stable frequency comb operations for both lasers, we operate the two lasers at a drive current around 790 mA as marked in Figures \ref{fr}a, \ref{fr}b, and \ref{fr}c. It is worth noting that the device dimensions, i.e., ridge width, cavity length, substrate thickness, etc., can influence the group velocity dispersions and finally affect the comb operation of the lasers. In Figure S1 in the Supporting Information, inter-mode beat note maps of the two other lasers with different cavity lengths, i.e., 5 and 6 mm, and identical ridge width (150 $\mu$m) and substrate thickness (100 $\mu$m), are shown for comparison. It can be seen that the cavity length of 5.5 mm is optimal for comb operation. The emission spectra of the two lasers operated at 790 mA were measured using a Fourier transform infrared spectrometer with a spectral resolution of 0.1 cm$^{-1}$. As shown in Figure \ref{fr}d, the two lasers show almost identical emission spectra, which is clearly ideal for the dual-comb operation. In Figure \ref{fr}e, we show the measured inter-mode beatnote and dual-comb spectra measured when the two lasers are operated at 790 mA and 32 K in free running. Two narrow inter-mode beatnotes, $f$$_{\rm{b1}}$=6.630 GHz and $f$$_{\rm{b2}}$=6.641 GHz, are clearly observed with a frequency difference of 11 MHz which determines the line spacing of the dual-comb spectra. Note that the values of $f$$_{\rm{b1}}$ and $f$$_{\rm{b2}}$ (hence their difference) are strongly dependent on the specific operating conditions of the two lasers, e.g., temperature, drive current, and optical coupling. Therefore, their values can be slightly different when the two lasers are working in different conditions. The dual-comb spectrum shown in Figure \ref{fr}e is centered around 4 GHz, and this frequency can be tuned by drive current (see Figure S2, Supporting Information). From the measured laser output power and the device mounting geometry, we can estimate that the power that is finally coupled into the detector Comb2 for the multiheterodyne process is around 140 nW \cite{LiACSPhoton}. It can be clearly seen that the two inter-mode beatnotes in Figure \ref{fr}e shows a significant difference in intensity. This is because we use Comb2 as the detector to measure the both signals. The terahertz light of Comb1 first propagates in the free space and then a part of the light (140 nW) reaches the detector (Comb2) for the heterodyne measurement. Therefore, the self-detected inter-mode beatnote $f$$_{\rm{b2}}$ is $>$40 dB stronger than $f$$_{\rm{b1}}$.

\section{Phase locking and stability evaluation}

In this work, we lock one of the dual-comb lines to a stable LO signal, which is implemented by sending the error signal to a PLL which dynamically controls the drive current of Comb2 (see Figure \ref{setup}a and the Experimental Section for the details of the PLL). The phase locking technique forces Comb2 to follow the behaviour of Comb1 and keeps the beating frequency of the two combs a constant for the modes that are close to the locked line. The dual-comb line that is selected for the phase locking results from the beating between two terahertz modes of the two laser combs. If we consider the same order ($m$) of the modes from the two laser combs, the frequencies of the two terahertz modes can be written as,
\begin{equation}\label{PLL}
f_{1}=f_{\rm{c}1}+mf_{\rm{b}1};
f_{2}=f_{\rm{c}2}+mf_{\rm{b}2},
\end{equation}
where $f_{\rm{c}1}$ and $f_{\rm{c}2}$ are the carrier offset frequencies of Comb1 and Comb2, respectively. Therefore, the frequency of the dual-comb line can be written as,
\begin{equation}\label{PLL}
f=f_{2}-f_{1}=f_{\rm{c}2}-f_{\rm{c}1}+m(f_{\rm{b}2}-f_{\rm{b}1}).
\end{equation}
In this work, when we lock $f$, we are actually forcing the sum of the carrier offset frequency difference ($f$$_{\rm{c}2}$-$f$$_{\rm{c}1}$) and the repetition frequency difference ($f$$_{\rm{b}2}$-$f$$_{\rm{b}1}$) to remain constant.

\begin{figure*}[!t]
	\centering
	\includegraphics[width=0.9\textwidth]{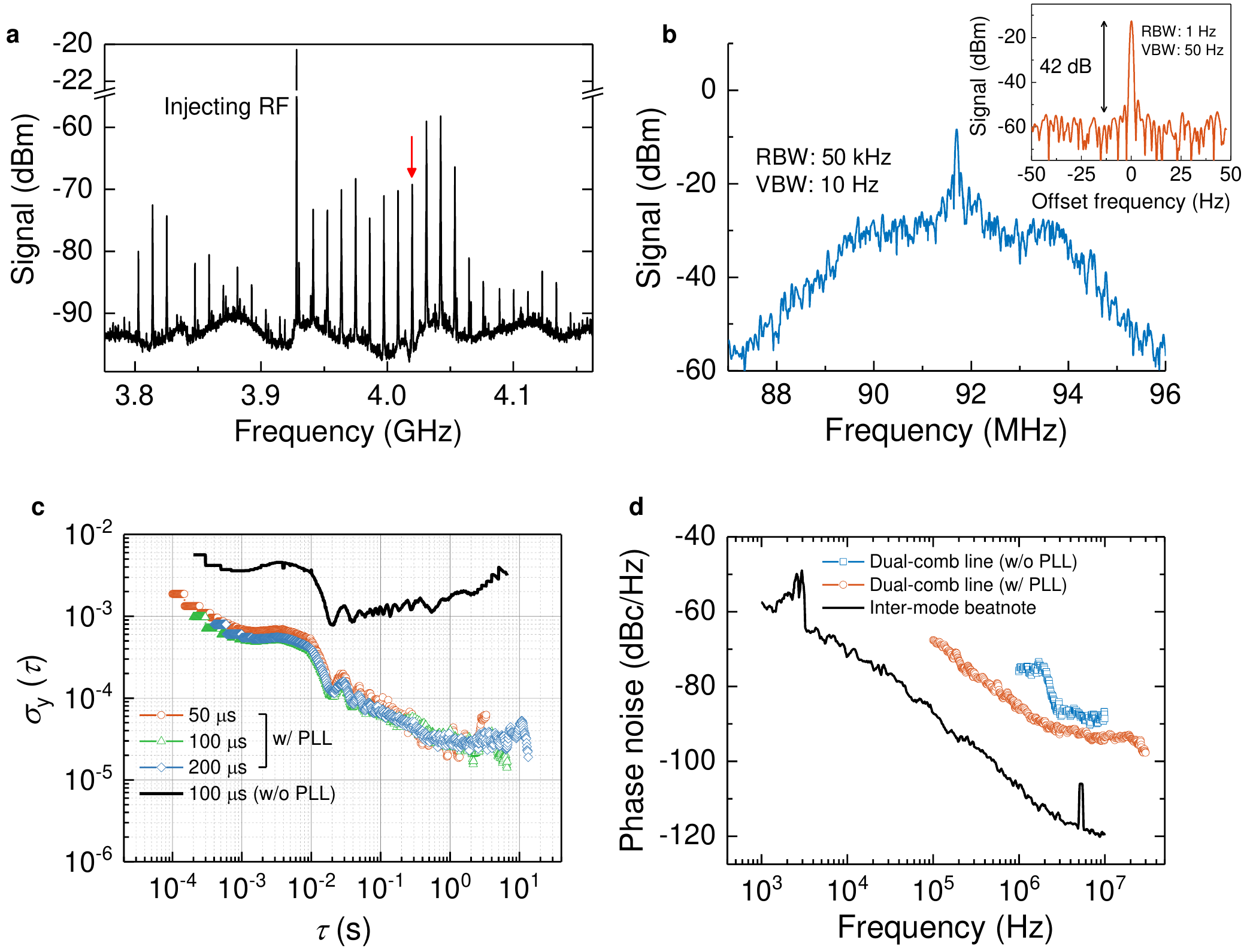}
	\caption{Phase locking of the terahertz dual-comb. a) Dual-comb spectrum around 4 GHz with one of the lines (indicated by the red arrow) phase locked to a stable RF synthesizer measured with a RBW of 100 kHz and a VBW of 500 Hz. The line with the highest power is the injecting RF that mixes the dual-comb line at 4.02 GHz (red arrow) down to 90 MHz. b) Phase locked line at 91.7079984 MHz measured with a RBW of 50 kHz and a VBW of 10 kHz. The inset is the high-resolution spectrum of the locked line measured with a RBW of 1 Hz. The offset frequency is subtracted by the central frequency of 91.7079984 MHz. c) Frequency Allan deviation plots measured using various gate times for the dual-comb line at 90 MHz with and without PLL. d) Phase noise spectra of the dual-comb line at 90 MHz with (circles) and without (squares) PLL. As a reference, the phase noise of the inter-mode beatnote signal at 6.64 GHz in free running is also shown as the solid line.}
	\label{locked}
\end{figure*}

{\textbf{Figure \ref{locked}}}a shows the dual-comb spectrum measured when one of the dual-comb lines at 4.02 GHz (marked by a red arrow) is phase locked. The strong signal at 3.9281 GHz is the injecting RF signal used to down-convert the dual-comb signal to 90 MHz. Then one of the down-converted dual-comb lines can be selected for phase locking. The spectrum of the locked line at 91.7079984 MHz is shown in Figure \ref{locked}b. The flat pedestal from 89.4 to 93.8 MHz shows the bandwidth of the bandpass filter. The inset is the high-resolution spectrum of the signal measured using a RBW of 1 Hz. It can be clearly seen that the line is phase locked, yielding a signal to noise ratio (SNR) $>$40 dB in 1 Hz bandwidth. In Video 1 in the Supporting Information, the phase locking process is recorded, and the effect on the line frequency stability with and without PLL can be clearly observed. The error signal vs. time that is sent to the Servo electronics for phase locking is also simultaneously recorded as shown in Video 2 in the Supporting Information.

To further quantify the frequency stability of the phase locked lines, we measured their Allan deviation and phase noise. In Figure \ref{locked}c, we report the frequency Allan deviation as a function of integration time, obtained using a frequency counter with various gate times (see Figure \ref{setup}a and the Experimental Section). For a clear comparison, the Allan deviation of the line at 90 MHz without PLL is also plotted (black solid curve). We can clearly see that the frequency stability is significantly improved once the phase locking is switched on. At short integration times around 1 millisecond, the calculated Allan deviation of the phase locked line is 6$\times$10$^{-4}$, which is approximately one order of magnitude lower than the value measured from the same line without PLL. More importantly, the results shown in Figure \ref{locked}c indicate that the phase locking improves the long-term frequency stability. As the integration time is longer than 0.04 s, the Allan deviation of the non-phase-locked line starts to increase, while it continues to decrease with time for the phase locked line. At an integration time of 1 s, the phase locked line shows an Allan deviation of 2.5$\times$10$^{-5}$ which is almost two orders of magnitude lower than the line without PLL. In Figure S3 in the Supporting Information, we also show the Allan deviation of the amplitudes of dual-comb lines in different situations. Similar to the frequency Allan deviations shown in Figure \ref{locked}c, the phase locked dual-comb line has a smaller amplitude Allan deviation than the dual-comb line without PLL.

\begin{figure*}[!t]
	\centering
	\includegraphics[width=0.9\textwidth]{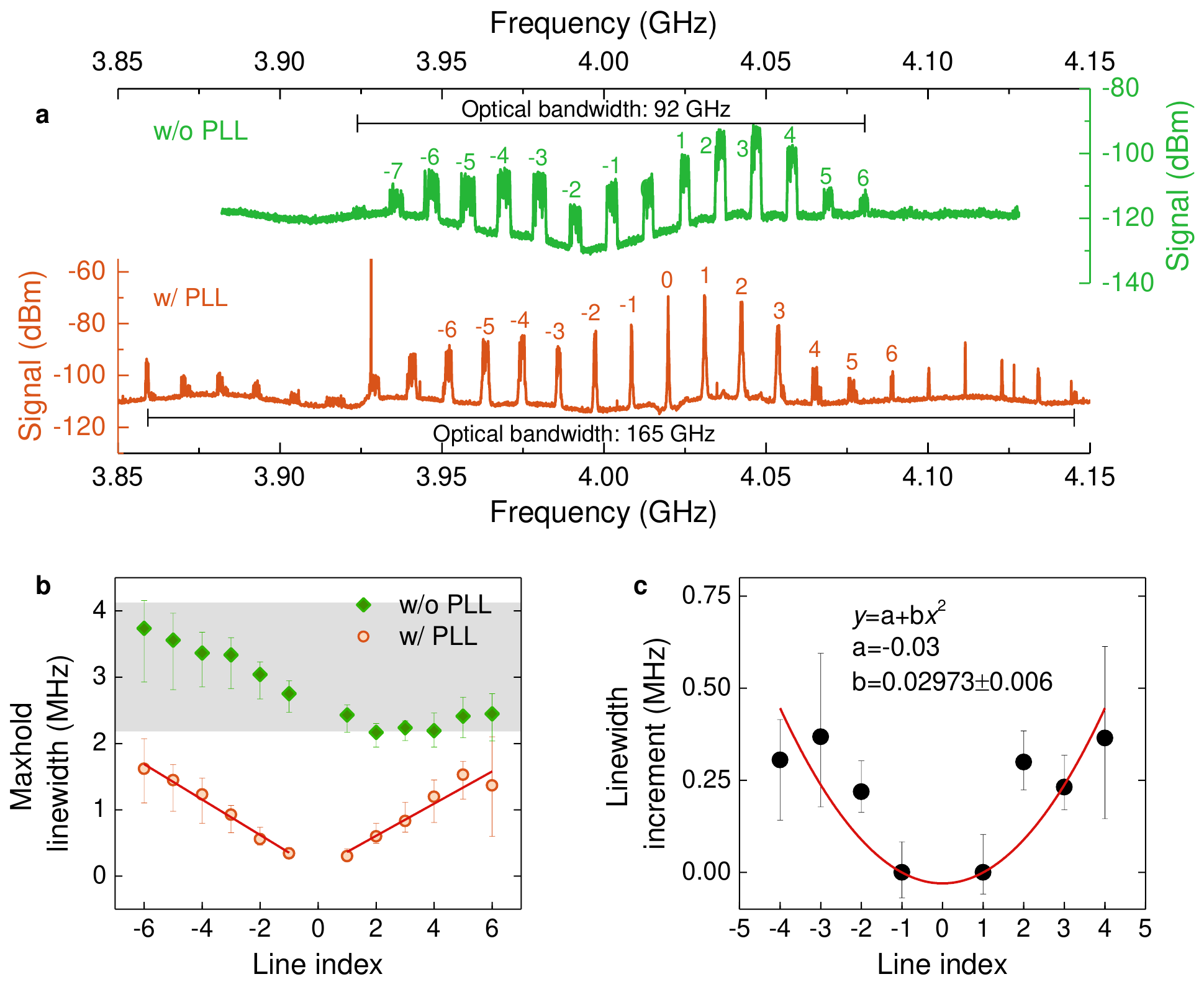}
	\caption{``Maxhold" measurement and linewidth comparison. a) Dual-comb spectra measured using the ``maxhold" function of the spectrum analyser. The measurement duration for each ``maxhold" trace acquisition is 30 seconds. The RBW and VBW parameters are set to 500 Hz and 50 Hz, respectively. b) Extracted ``maxhold" linewidth as a function of line index for the dual-comb spectra without (green diamonds) and with (red circles) PLL. The line indices shown in b are identical to the ones shown in a. Red lines are a guide to the eye. c) Linewidth increments for the closest 4 dual-comb lines with respect to the locked line extracted from b. The red curve in c shows the expected linewidth increment as a function of line index $n$, proportional to $n^2$. The error bars shown in b and c were obtained through multiple measurements. The dots show the average values of three consecutive measurements and the two ends of each error bar indicate the minimum and maximum values reached with all measurements.}
	\label{maxhold}
\end{figure*}

Figure \ref{locked}d plots the phase noise spectra for the dual-comb lines at 90 MHz with and without PLL, which were measured using a phase noise module (K40) of the spectrum analyser (Rohde \& Schwarz, FSW 26). The RBW parameters used for the phase noise measurement are scaled with the frequency offset from the carrier. At any frequency, the phase locked line always shows a lower phase noise than the dual-comb line without PLL. At 100 kHz and 1 MHz, the measured phase noises of the dual-comb line with PLL are -67 and -86 dBc/Hz, respectively. In Figure \ref{locked}d, the phase noise spectrum of the inter-mode beatnote signal of Comb2 in free running is also plotted, showing a much smaller phase noise than the dual-comb lines with and without PLL. It is measured that the phase noise is -106 dBc/Hz (-87 dBc/Hz) at 1 MHz (100 kHz) for the free running inter-mode beatnote signal at 6.64 GHz. The stable inter-mode beatnote finally contributes to the dual-comb operation and the phase locking of the dual-comb lines.

To directly show the effect of the phase locking on the long-term frequency stability of the dual-comb spectra, in Figure \ref{maxhold}a we plot the dual-comb traces without (green) and with (red) PLL acquired with the ``maxhold" function of the spectrum analyser for 30 seconds. We can clearly observe a significant improvement of the long-term frequency stability for all the dual-comb lines once the phase locking is switched on. Furthermore, the number of dual-comb lines is increased from 15 without PLL to 25 with PLL, i.e., the corresponding optical bandwidth is increased from 92 to 165 GHz as indicated in Figure \ref{maxhold}a. The measured ``maxhold" linewidths for the closest lines with respect to the phase locked line are plotted in Figure \ref{maxhold}b. The line marked by ``0" is the phase locked line. It can be clearly seen that as the lines are farther away from the phase locked line, the measured ``maxhold" linewidths increase with the line index. The measured ``maxhold" linewidths for the closest 6 dual-comb lines are smaller than 2 MHz with a smallest linewidth of 220 kHz for line 1. For comparison, the ``maxhold" linewidths of the dual-comb lines without PLL extracted from the green curve of Figure \ref{maxhold}a are also shown as diamonds in Figure \ref{maxhold}b. As shown by the shaded area in Figure \ref{maxhold}b, the ``maxhold" linewidths for dual-comb lines without PLL span a range between 2 and 4 MHz. In Figure \ref{maxhold}c, we plot the linewidth increment as a function of line index for the 8 closest dual-comb lines (4 on each side). When the frequency of a signal is multiplied by $n$, then its phase noise, hence its linewidth, is multiplied by $n^2$. Therefore, the expected linewidth increment in Figure \ref{maxhold}c is proportional to $n^2$, since, moving away from the line of index ``0", each line oscillates at an integer multiple of ($f$$_{\rm{b}2}$-$f$$_{\rm{b}1}$). The red curve in Figure \ref{maxhold}c is a fit to the experimental data. For $n$=-1 and 1, the linewidth increment is set as 0. Therefore, an intercept should be added into the fitting function. The fitting function can be written as $y$=a+b$x$$^2$, where a is the intercept, b is the fitting parameter. To satisfy $y$=0 when $x$=-1 and 1, a is fixed at -0.03. The fitting gives a value of 0.02973 for b with a standard deviation of 0.006.

The experimental results shown in Figures \ref{locked} and \ref{maxhold} prove that the dual-comb lines can be phase locked using the traditional PLL technique. We lock one of the dual-comb lines, $f$ (the sum of the carrier offset frequency difference and repetition frequency difference), to improve the stability of the dual-comb laser source. Since the two repetition rates are not locked, we, in principle, perform the partial phase locking of the dual-comb source and improve the stabilities of dual-comb lines that are close to the phase locked line. Although the technique is not a complete locking of the dual-comb source, it works with majority of dual-comb lines. Actually, two prerequisites should be satisfied for a successful locking. The first one is the power requirement. To mix with the LO signal (RF generator 2 in Figure \ref{setup}a) and generate the error signal for the phase locking, the power of the selected down-converted dual-comb line should be greater than -30 dBm (after microwave amplifications). The second condition is that the instantaneous linewidth of the signal should be smaller than the feedback bandwidth of the PLL (400 kHz). In this work, the majority of the dual-comb lines satisfy both conditions and can be used to implement the phase locking. 
		
There are potential ways to further improve the effectiveness of the locking. First of all, the complete locking can be further implemented. At present, we locked $f$ rather than the repetition rates and/or carrier offset frequencies. If the repetition rates of the two lasers can be locked simultaneously using the RF injection locking technique \cite{GellieOE,LiOE}, the carrier offset frequency difference ($f$$_{\rm{c}2}$-$f$$_{\rm{c}1}$) is then locked. Afterwards, the complete locking of the dual-comb laser source is achieved. Secondly, the optical coupling and thermal management are crucial for the comb stabilities. For example, the distance between the two laser combs can be optimized or a lens coupler can be used to tune the optical coupling for an ideal dual-comb operation; the sample holder can be re-designed to improve the thermal management for a more stable comb operation. Once the power and stabilities of the comb and dual-comb signals are improved, the locking will be more effective.

\section{Time traces of the dual-comb signal}

\begin{figure*}[!t]
	\centering
	\includegraphics[width=0.9\textwidth]{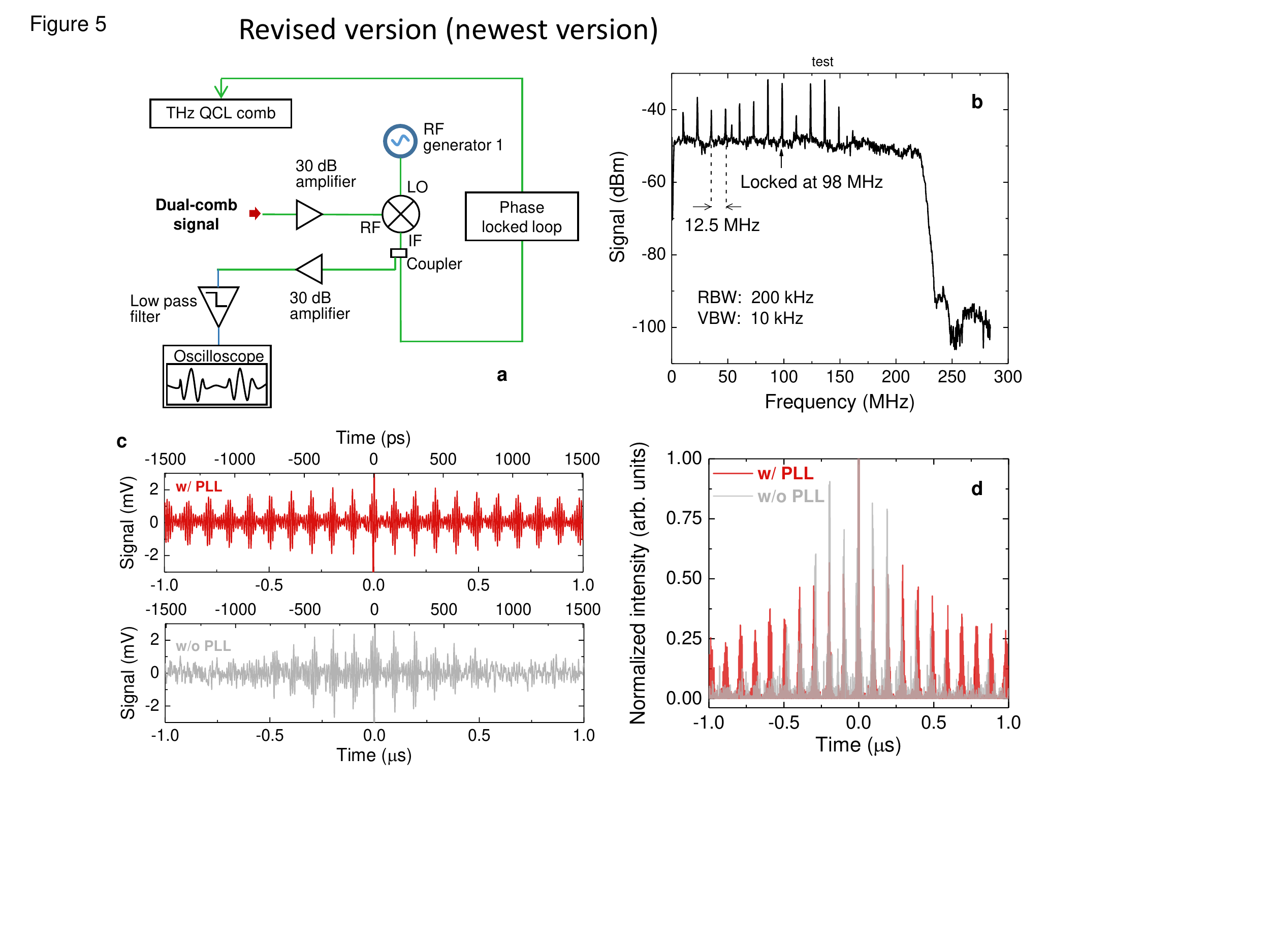}
	\caption{Time trace measurement of the terahertz dual-comb source. a) Experimental setup of the time trace measurement of the dual-comb signal. b) Down-converted dual-comb spectrum measured with a low pass filter with a cutoff frequency of 200 MHz. The dual-comb line at 98 MHz is phase locked. A RBW of 200 kHz and a VBW of 10 kHz were used for the dual-comb spectrum measurement. c) Measured time traces of the down-converted and filtered dual-comb signal with (upper panel) and without (lower panel) PLL. In the top x-axes of both panels, the estimated time scale of terahertz pulses are showed by considering the repetition frequencies of single combs (see Figure \ref{fr}e). d) Normalized intensity plots of the data in c. All measurements are performed when QCL comb1 and comb2 are operated at 795 mA and 778 mA, respectively, at a stabilized temperature of 31 K.}
	\label{time}
\end{figure*}

We also evaluate the time characteristics of the dual-comb signal. {\textbf{Figure \ref{time}}}a shows the experimental setup used for the time trace measurement of the dual-comb signal. As in Figure \ref{setup}a, the dual-comb signal is first down-converted close to 90 MHz, then the signal is split using a microwave coupler: one part is used for phase locking and the other to record the time trace. To obtain a clean signal for the time domain measurement, a low pass filter with a cutoff frequency of 200 MHz is used before the signal is sent to an oscilloscope. The dual-comb signal is shown in Figure \ref{time}b, where the line at 98 MHz is phase locked to a stable RF synthesizer. The time traces with and without PLL are shown in Figure \ref{time}c on a time window of 2 $\mu$s. It can be clearly seen that in the entire time domain, the dual-comb signal with PLL shows a periodic pulse train, while for the dual-comb signal without PLL, pulses can be only observed in a limited time range of 1 $\mu$s, with a pulse shape changing from pulse to pulse. Beyond $\sim$0.5 $\mu$s-delay no pulse can be observed. It is worth noting that the high peaks at the center of the horizontal axis (0 $\mu$s delay) are artificially introduced by the use of the low pass filter. As shown in Figure \ref{time}b, the low pass filter introduces a sharp cutoff around 200 MHz in the RF spectrum, which results in a strong peak at 0 delay in the time trace after a Fourier transform (see Figures S4 and S5, Supporting Information). Figure \ref{time}d plots the intensity of the time traces with and without PLL. One can see the signal with PLL shows a fast decay within 1 $\mu$s because the pulses lose the coherence, which is in agreement with the measured ``maxhold" linewidths of $\sim$1 MHz shown in Figure \ref{maxhold}b. Without PLL, the observed intensity decay goes faster, which again agrees well with the larger linewidths as shown in Figure \ref{maxhold}b for the case without PLL.

It is worth noting that the phase locking technique employed in this work is applied to the dual-comb signal to force one of the laser combs to follow the behaviour of the other laser. The target of the phase locking is to improve the frequency stability of the dual-comb operation. As shown in Figure \ref{setup}a, we don’t use any modulation and/or phase locking technique to control the repetition rates of the two lasers. Although the dual-comb time trace shown in Figure \ref{time}c does not provide the output waveform of one single QCL, our result indicates a clear time periodicity with a pulsed-type behaviour. Actually, the technique can be viewed as an asynchronous sampling, as shown in Ref. \cite{BarbieriNP} where a terahertz QCL comb was phase locked to an ultra-stable femtosecond laser and the comb lines were down-converted to low frequencies for time trace measurements. Here, we use one QCL comb as ``LO" for down-conversion: due to the high frequency stability and coherence of the two QCL combs, the comb lines can be also down-converted for time domain measurements. More importantly, in Ref. \cite{BarbieriNP} it is a must to apply the active mode locking onto a terahertz QCL to obtain stable pulses, while in this work we show that the free running terahertz QCL can emit pulsed-type waveforms. 

\section{Conclusion}
In summary, we have demonstrated the active stabilization of a terahertz QCL dual-comb source emitting around 4.2 THz using a traditional phase locking technique. By phase locking one of the down-converted dual-comb lines to a RF synthesizer, we are able to demonstrate the partial mutual phase lock of the two combs, and finally significantly improve the frequency stability or coherence of dual-comb lines that are close to the phase locked line. Experimental results reveal that the phase locked dual-comb lines show much improved frequency/amplitude Allan deviations, phase noise, and “maxhold” linewidths than the dual-comb lines without PLL. The achieved stability could be further improved by injection locking of the two combs repetition rates using external RF synthesizers \cite{GellieOE}. Finally, we experimentally show that the dual-comb signal with PLL can produce periodic time pulses, which indicates that the terahertz QCL comb without a control of the repetition rate is emitting pulsed-type waveforms.


\section*{Experimental Section}

\subsection{Terahertz QCL}

The active region of the terahertz QCL used in this work is based on a hybrid design where bound-to-continuum transitions are exploited for the generation of terahertz photons, while fast longitudinal optical phonon scattering allows the fast depopulation of the lower laser state. The QC structure is designed for emission at 4.2 THz and the detailed layer structure can be found in ref. \cite{WanSR}. The QCL active region was grown by a molecular beam epitaxy system on a semi-insulating GaAs (100) substrate. The grown wafer was then processed into a single plasmon waveguide geometry with a ridge width of 150 $\mu$m. To improve the thermal management, the laser substrate was thinned down to 100 $\mu$m using grinding and polishing techniques. Various cavity lengths (between 5 and 7 mm) were obtained by direct cleaving of the fabricated laser ridges. We finally selected a cavity length of 5.5 mm in this work because it experimentally allowed optimal frequency comb operation (see Figure \ref{fr} and Figure S1, Supporting Information). For the laser dual-comb characterizations, two terahertz QCLs with identical cavity length (5.5 mm) and ridge width (150 $\mu$m) were mounted onto a ``Y" shape cold finger, see the inset of Figure \ref{setup}a.

The cw output power of the terahertz QCLs was measured using a terahertz power meter (Ophir, 3A-P THz) and the lasers were operated in constant current mode. To measure the power as accurately as possible, two off-axis parabolic mirrors were used for collecting and collimating the terahertz light into the power sensor. In addition, the beam path was purged with dry air to reduce water absorption. The emission spectra shown in Figure \ref{fr}c were measured with a Fourier transform infrared (FTIR) spectrometer (Bruker, v80) with a spectral resolution of 0.1 cm$^{-1}$ (3 GHz).

\subsection{Laser self-detection}

Because carrier inter-subband transitions in QCLs are fast processes unfolding on a time scale of a few picoseconds, the QCL itself has an ultrafast response time when perturbed from equilibrium\cite{LiOE}. When a modulated terahertz beam is incident onto a QCL detector, the population inversion will be modulated, giving rise to a current modulation that can be measured directly using a spectrum analyser. As a result, for the RF measurements carried out in this work, the QCL can be inherently used as a fast detector as well as a terahertz emitter\cite{GellieOE,THzDualComb,LiPRA,LiACSPhoton}. Both the inter-mode beat note and down-converted dual-comb spectra can be measured simultaneously employing such laser self-detection scheme and recorded using a spectrum analyser (Rohde \& Schwarz, FSW26). 

As shown in Figure \ref{setup}a, in this experiment Comb2 is used as a fast detector. The detected heterodyne signal is sent to a Bias-T (Marki, BT2-0026) and then amplified for different measurements. To evaluate the long-term frequency drift, the ``maxhold" measurement is used as shown in Figure \ref{maxhold}a. When the ``maxhold" function of the spectrum analyser is switched on, for a given measurement time duration the spectral maxima are recorded for each temporal sweep.

\subsection{Phase locked loop (PLL)}

Active phase locking has been used in the past to lock the repetition rate of a single QCL comb and/or lock a QCL line to a stable source\cite{BarbieriNP,ConsolinoNC2019,CappelliNP}. Here, we use the PLL to lock one terahertz QCL dual-comb line. A commercial product (ppqSense S.r.l., QubeCL) has been used to perform the phase locking experiment. It consists of a low noise current source, used to drive Comb2, and a phase lock module. When the PLL is activated, a current proportional to the phase error signal is processed by a Proportional-Integral (PI) stage and then it is internally added to the bias current. The total current is injected into Comb2 through a Bias-T (see Figure \ref{setup}a) to dynamically control its emission. The feedback bandwidth of the PLL is 400 kHz. The experimental setup is shown in Figure \ref{setup}a. Different from the repetition rate locking of a single QCL comb where the inter-mode beatnote signal is sent to a phase lock loop, here one of the dual-comb lines is filtered out for the phase locking. Next, the detected heterodyne signal is sent to a Bias-T for RF measurements. The output of the AC port of the bias-T carries a signal carrying different spectral informations, including the inter-mode beatnotes and their harmonics, the dual-comb spectra at different carrier offset frequencies, and also the reflected signals from the RF generators (local oscillators). The signal is first amplified and mixed with RF generator 1. To down-convert the dual-comb spectrum, the frequency of RF generator 1 is fine tuned to finally obtain a series of dual-comb lines around 90 MHz. Then, a tunable bandpass filter with a bandwidth of 5 MHz is used to select just one dual-comb line. The selected line is amplified and sent to a mixer to beat with the signal from RF generator 2 (90 MHz). Finally, the error signal is sent to a Servo electronics to dynamically control the drive current of Comb2. For a stable locking, the power values of RF generator 1 and RF generator 2 are set to be 10 dBm and 0 dBm, respectively.

\subsection{Allan deviation measurement}

The frequency Allan deviations are measured using a frequency counter with a bandwidth of 350 MHz (Keysight, 53230 A). To accurately measure the single frequency signal, a filter with a bandwidth of 5 MHz is used to filter out other frequency components. The frequency counter is set to work in a continuous mode and the sampling rate is determined by setting the gate time parameter.

The recorded frequency values as a function of time can be used to calculate the Allan deviation for various integration times $\tau$ by using the following equation
\begin{equation}
\sigma_{\rm{y}}(\tau)=\left[\frac{1}{2M}\Sigma_{i=0}^{M-1}(y(i+1)-y(i))^2\right]^{1/2},
\end{equation}
where $M$=$T$/$\tau$-1 with $T$ being the entire measurement interval, $y(i)$=$\frac{<\alpha(t_0+i\tau)>_\tau-\alpha_0}{\alpha_0}$ with $<\alpha(t_0+i\tau)>_\tau$ being the frequency averaged over a time interval $\tau$ and $\alpha_0$ the mean value of the measured frequencies.

\subsection{Time trace measurement}

To obtain the time characteristics of the pure dual-comb signal, we first down-convert the latter to lower frequencies and filter out unwanted frequency components (i.e., higher harmonics, etc.) by using a low pass filter with a nominal cutoff frequency of 200 MHz as shown in Figure \ref{time}a. The time traces shown in Figure \ref{time}c are measured using an oscilloscope with a bandwidth of 20 GHz. Before filtering the down-converted dual-comb signal, the latter is amplified by 30 dB. The waveform is recorded using 2048 averages.


\section*{References}
\bibliographystyle{naturemag}
\bibliography{REF}


\begin{addendum}
 \item This work is supported by the National Natural Science Foundation of China (61875220, 62035005, 61927813, 61704181, and 61991432), the “From 0 to 1” Innovation Program of the Chinese Academy of Sciences (ZDBS-LY-JSC009), the Major National Development Project of Scientific Instrument and Equipment (2017YFF0106302), the Scientific Instrument and Equipment Development Project of the Chinese Academy of Sciences (YJKYYQ20200032), National Science Fund for Excellent Young Scholars (62022084), Shanghai Outstanding Academic Leaders Plan (20XD1424700), and Shanghai Youth Top Talent Support Program. The authors thank R. Qian and X. W. Sun for a loan of a RF synthesizer.

 \item[Author contributions] H.L. conceived the study. Z.P.L., W.J.W., Y.R.Z., W.G., and S.J.Y. fabricated the terahertz QCLs and carried out the basic device measurement. Y.R.Z., Z.P.L., K.Z., X.Y.L., W.G., and H.L. performed dual-comb measurements. K.Z. and H.L. did the noise analysis. Y.R.Z., Z.P.L., X.Y.L., W.G., and H.L. performed the phase locking of the terahertz dual-comb source. H.L., S.B., Y.R.Z., Z.P.L., K.Z., J.C.C, and D.X. participated in the date analysis. H.L. wrote the manuscript with contributions from all other authors. H.L. supervised the project.

 \item[Addition information] Correspondence and requests for materials should be addressed to H.L. (email: hua.li@mail.sim.ac.cn).

 \item[Competing financial Interests] The authors declare that they have no competing financial interests.

\end{addendum}

\end{document}